\documentclass[showpacs, amsmath, amssymb, amsthm, aps, jmp, thmsb, twocolumn, superscriptaddress]{ptephy_v1}

\usepackage{color}
\usepackage{float}
\usepackage{amsmath}
\usepackage{amssymb}
\usepackage{amsfonts}
\usepackage{amsbsy}
\usepackage{latexsym}
\usepackage{mathrsfs}
\usepackage{bbm}
\usepackage{bm}

\newtheorem{proposition}{Proposition}
\newtheorem{theorem}{Theorem}
\newtheorem{lemma}{Lemma}
\newtheorem{corollary}{Corollary}
\newtheorem{assumption}{Assumption}

\newcommand{\be}{\begin{eqnarray}}
\newcommand{\ee}{\end{eqnarray}}
\def\({\left(}
\def\){\right)}
\def\[{\left[}
\def\]{\right]}
\def\C{\mathbb{C}}
\def\R{\mathbb{R}}

\newcommand{\braket}[1]{\left\langle #1 \right\rangle}

\newcommand{\bra}[1]{\langle #1 |}
\newcommand{\ket}[1]{| #1 \rangle}

\newcommand{\sla}[1]{\rlap{\kern .15em /}#1}

\newcommand{\pro}{\mathrm{Pr}}

\newcommand{\vz}{\bm{z}}

\newcommand{\tr}{{\rm tr}}
\newcommand{\vpsi}{\bm{\psi}}
\newcommand{\vphi}{\bm{\varphi}}
\newcommand{\vxi}{\bm{\xi}}

\begin{document}
\title{Born Rule and Logical Inference in Quantum Mechanics}

\author{Tsubasa Ichikawa}
\affil{Mukogaoka 1-10-6-1004, Tokyo 113-0023, Japan}

\begin{abstract}
Logical inference leads to one of the major interpretations 
of probability theory called logical interpretation, in which the probability is seen 
as a measure of the plausibility of a logical statement 
under incomplete information. In this paper, assuming that our
usual inference procedure makes sense for every set of logical propositions 
represented in terms of commuting projectors on a given Hilbert space, 
we extend the logical interpretation to quantum mechanics and derive the Born rule. 
Our result implies that, from the epistemological viewpoints, we can regard quantum mechanics as a natural extension of the classical probability theory.
\end{abstract}
%\pacs{%03.65.Vf, 03.67.Pp, 82.56.Jn.
%}

\maketitle

%%%%%%%%%%%%%%%%%%%%%%%%%%%%%%
\section{Introduction}
%%%%%%%%%%%%%%%%%%%%%%%%%%%%%%
Inference is one of the essential building blocks in various branches
of mathematical science: Jaynes' derivation of statistical mechanics \cite{Jaynes57a, Jaynes57b} is based on the inference of the probability
distribution which is most likely to reproduce given expectations of thermodynamic variables.
Since the era of Laplace \cite{Laplace02}, probability has been more or less
regarded as a measure of likelihood of statements being valid,
although there coexists the interpretation that the probability is nothing but
the relative frequency.

The inference plays an important role even to understand fundamental 
aspects of probability theory. The following is shown by Cox \cite{Cox46, Cox61, Halpern99a, Halpern99b, Jaynes03, vanHorn03}: Suppose there is a measure of the plausibility
of statements being valid under incomplete information. If the measure
follows inference rules consistent with our common sense,
then the measure satisfies the product rule and 
the sum rule of the probability, and hence it can be interpreted as the probability. 
In other words, without invoking the  notion of the
relative frequency, probability theory can be derived by
assuming reasonable inference rules  under incomplete information.
This derivation is called Cox's theorem, which allows us to 
see the probability as a natural extension of the truth values
of the propositions. This interpretation of probability
is called the logical interpretation of probability \cite{Kaynes21}.

Recently, several inference methods are becoming utilized
to reconstruct quantum mechanics. The equations of motions
such as the Schr\"odinger equation \cite{Raedt14}, the Pauli equation \cite{Raedt15}, 
and the Klein-Gordon
equation \cite{Donker16} are derived under the logical interpretation, augumented with the auxiliary requirement that
the plausibility of the experimental outcomes is robust under the slight changes
of the experimental parameters. Wave function is also derived with other auxiliary assumptions \cite{Caticha98, Caticha14}.

In this paper, we show that the measure of the plausibility takes the form of the Born rule by extending the argument mentioned above to the Hilbert space and the projectors thereof.
Interestingly enough, the inference rule leads to the unique concrete expression of the quantum conditional probability associated with the L\"uders conditionalization \cite{Luders50}, which specifies the post-measurement quantum state. 
More precisely, the L\"uders conditionalization requires that after a projection measurement $Q$ on a density matrix $\rho$, the state changes to $\rho_Q=Q\rho Q/\tr(\rho Q)$, whereas we will find that the quantum conditional probability in accordance with the inference rule takes the form $\pro(P|Q)=\tr(\rho_QP)=\tr(Q\rho QP)/\tr(\rho Q)$, where $P$ is a projection operator. Thus the quantum conditional probability is seen as the Born rule for the density matrix $\rho_Q$, which is obtained after the measurement of $Q$. We hereafter call the expression of the quantum conditional probability L\"uders rule, in order to stress on the relationship between the L\"uders conditionalization and the quantum conditional probability \cite{Busch09}.

Analogously to the derivation of Cox\rq s theorem, throughout our derivation of both the rules, we will assume that the plausibility measures follow our standard inference procedures for commuting projectors.
Our derivation, hence, enables us to see both the rules as the measures of plausibility on which our inference makes sense for the commuting projectors.

This paper is organized as follows. In Section~\ref{formulations},
we review two formulations of the probability theory. 
In Section~\ref{interpretations}, we present three
approaches to interpretations of the probability theory. One of those, Cox's approach to the logical interpretation on the basis of inference
rules, is studied in detail. In Section~\ref{extension}, we extend Cox's approach in quantum mechanics.
In Section~\ref{Lueders}, we characterize the L\"uders rule as the unique conditional probability satisfying the inference rule.
In Section~\ref{violation}, we give an example that the inference rule employed in Cox's approach is no longer valid for non-commuting projectors.
Section~\ref{conclusion} is devoted to our conclusion and discussions.

%%%%%%%%%%%%%%%%%%%%%%%%%%%%%%
\section{Two Formulations of Probability Theory}
\label{formulations}
%%%%%%%%%%%%%%%%%%%%%%%%%%%%%%

Probability and conditional probability are central notions in the probability theory. Depending upon which one we take as the more fundamental concept, we have two formulations of the probability theory.
In this section, we give a brief review of them, since we need both in the following argument. 

We hereafter deal with both the formulations based on the finite additive class, which is defined as follows.
Suppose we are given countable infinite number of mutually different elements $\omega_1, \omega_2, \cdots$. 
The set of all the elements is denoted by $\Omega:= \{ \omega_1, \omega_2, \cdots \} $, and the subset thereof is written by $A_1, A_2, \cdots$. 
A set ${\cal F}$ of the subsets $A_1, A_2, \cdots, A_n$ is called the finitely additive class
if it satisfies
\be
\emptyset&\in&{\cal F},\nonumber\\
A_i&\in&{\cal F} \quad \Rightarrow \quad A_i{}^c\in{\cal F},\nonumber\\
A_1, A_2, \cdots A_n&\in&{\cal F} \quad \Rightarrow \quad \bigcup_{i=1}^n A_i\in{\cal F}.\label{defF}
\ee
Here $\emptyset$ stands for the empty set and $A{}^c$ is the complement of $A$ defined through $A\cap A{}^c=\emptyset$ and $A\cup A{}^c=\Omega$. 
The family of the sets $\{A_i\}_{i=1}^\infty$ satisfying Eqs.~(\ref{defF}) for $n\rightarrow\infty$ is called the $\sigma$-algebra.

Let us turn to the two formulations mentioned above. In the first one, which we hereafter call the probability-based formulation, the probability is given first, and the conditional probability is defined as the ratio thereof. 
%We first recall the standard definition of probability.
More precisely, the probability $\pro(A)$ is defined  as a map from ${\cal F}$ to $[0,1]$, which satisfies 
\begin{subequations}
\begin{align}
\pro(A)&\ge0\quad\text{for}\quad \forall A\in {\cal F}, \label{null}\\
\pro(\Omega)&=1,\label{omega}\\
\pro(\bigcup_{i=1}^n A_i)&=\sum_{i=1}^n\pro(A_i),
%\quad\text{for}\quad\{A_i\,|\,A_i\cap A_j=\emptyset \forall i, j \in\N\}.
\label{ad}
\end{align}
\label{Kol}
\end{subequations}
if $\{A_i\}_{i=1}^n$ is a family of mutually disjoint subsets of $\Omega$: $A_i\cap A_j=\emptyset$ for $i\neq j$.
The conditional probability $\pro(A|B)$ is given as
\be
\pro(A|B)=\frac{\pro(A\cap B)}{\pro(B)},
\label{ratio}
\ee
when $\pro(B)\neq0$.

The second formulation \cite{Renyi55, Renyi56, Renyi70}, in contrast, is based on the conditional probability. In this formulation, we introduce the finite additive class ${\cal F}$ and its non-empty subset ${\cal G}$, and define the conditional probability $\pro(A|B)$ for $A\in{\cal F}$ and $B\in{\cal G}$ as the non-negative map, which fulfills
\begin{subequations}
%\begin{align}
\begin{align}
	\pro(B|B)&=1\quad\text{for}\quad \forall B\in {\cal G}, \label{one}\\
	\pro(\bigcup_{i=1}^n A_i|B)&= \sum_{i=1}^n\pro(A|B) \label{c-additivity}
\end{align}
for the set $\{A_i\}_{i=1}^n$ of the mutually disjoint elements and $B\in{\cal G}$, and
\begin{equation}
	\pro(A|B)=\frac{\pro(A\cap B|C)}{\pro(B|C)}
\label{abc}
\end{equation}
%\end{align}r
\label{Renyi}
\end{subequations}
for $A\in{\cal F}$ and $B,C \in{\cal G} $ such that $B\subset C$ and $\pro(B|C)>0$.

The formulation based on Eqs.~(\ref{Renyi}) is a natural extension of the probability-based formulation. It is shown in \cite{Renyi56} that the conditional probability $\pro(A|B)$ in the sense of Eq.(\ref{Renyi}) can be seen as the probability when $B$ is fixed. Thus, we utilize the notation $\pro(A)$ to the special case $B=\Omega$, that is,
\be
\pro(A)=\pro(A|\Omega).
\ee 
Conversely, the conditional probability (\ref{ratio}) satisfies Eqs.~(\ref{Renyi}). Note that not all the conditional probabilities in the sense of Eqs.~(\ref{Renyi}) take the form of Eq.~(\ref{ratio}). 
These two formulations are extended to the $\sigma$-algebra by setting $n\rightarrow\infty$ in Eqs.~(\ref{Kol}) and (\ref{Renyi}), respectively.

%%%%%%%%%%%%%%%%%%%%%%%%%%%%%%
\section{Interpretations of Probability}
\label{interpretations}
%%%%%%%%%%%%%%%%%%%%%%%%%%%%%%

To give the interpretations of the probability theory,
we construct a mathematical quantity 
which is easy to understand intuitively and 
fulfills the conditions (\ref{Kol}) or (\ref{Renyi}), depending on which formulation we take.
Here we take three examples of well-known
interpretations of the probability, that is, frequency interpretation, subjective interpretation, and logical interpretation \cite{Gillies00}.
Note that the frequency interpretation and subjective interpretation are applicable to the finitely additive class as well as $\sigma$-algebra \cite{Williamson99}. 
On the other hand, the logical probability is shown to be applicable to the finitely additive class.

%%%%%%%%%%%%%%%%%%%%%%%%%%%%%%
\paragraph*{Frequency Interpretation.}
%%%%%%%%%%%%%%%%%%%%%%%%%%%%%%
First, let us show that the relative frequency 
of the events occurring satisfies the conditions (\ref{Kol}). 
Consider a coin flipping game, whose possible outcomes
are either finding the head (H) or the tail (T) of the coin.
Thus, we have $\Omega=\{\rm{H}, \rm{T}\}$ and ${\cal F}=\{\emptyset, \{\rm{H}\}, \{\rm{T}\}, \Omega\}$.  Note that $\Omega=\{\rm H\}\cup\{\rm T\}$ and $\{\rm H\}\cap\{\rm T\}=\emptyset$.
Now we repeat the coin flipping
game $N$ times and write the number of finding H as $N_{\rm H}$ and
that of finding T as $N_{\rm T}$, respectively. The relative frequency
of finding H is $N_{\rm H}/N$ and that of finding T is $N_{\rm T}/N$.
By setting $\pro(\emptyset)=0$, $\pro(\{{\rm H}\})=N_{\rm H}/N$,
$\pro(\{{\rm T}\})=N_{\rm T}/N$, and $\pro(\Omega)=1$, we find
\be
\pro(\Omega)=\pro(\{\rm H\}\cup\{\rm T\})=\pro(\{{\rm H}\})+\pro(\{{\rm T}\})=1,
\ee
which shows that the relative frequency satisfies the conditions (\ref{Kol}). 
This is the reason why we may interpret the probability as relative frequency.

In the frequency interpretation, we need a well-defined ensemble of the repeatable events
to define the probability distributions. On the other hand, it has been pointed out that 
the frequency interpretation is not applicable to one-shot event or the measurement of physical constants \cite{Cousins95, Lyons13}. The following two interpretations
provide useful tools for the analyses of such cases.

%%%%%%%%%%%%%%%%%%%%%%%%%%%%%%
\paragraph*{Subjective Interpretation.}
%%%%%%%%%%%%%%%%%%%%%%%%%%%%%%

The second example is  Dutch Book argument (DBA) \cite{deFinetti37}. In DBA,
we consider a bet on whether a given hypothesis (it rains tomorrow, for example) is true.
If the hypothesis is true, then the bettor obtains the stake $S$.
If it is not the case, he obtains nothing. Let us now suppose that the bettor pays
the wager $qS$, where $q$ is called the betting
quotient. Then the net payoff of the bet is $S-qS$ if the hypothesis is true, 
and $-qS$ otherwise.

In \cite{deFinetti37}, de Finetti showed that the bookie can construct a set of the bets in such a way that the bettor always get the net loss, if the set of the
betting quotients does not obey the conditions (\ref{Kol}). In other words, as far as the bettor is rational in the sense that he wish to avoid the net loss, his betting quotients have to obey the axioms of the probability theory: the fair betting quotients can be seen as the probability. See \cite{Gillies00, Caves02} for the proof on the mathematical relation between the fair betting quotients and the axioms of the probability.

DBA leads to the subjective interpretation of the probability, since
the betting quotients indicates how much the bettor feels the hypotheses true. 
The salient feature of DBA is that the bettor can freely determine his own betting quotients:
In a bet, it is not necessary that all the participants of the bets have the same betting quotients. This implies that in the subjective interpretation we may assign
several differenct probabilities (degree of belief) for the plausibility of a hypothesis by our own decision.

%%%%%%%%%%%%%%%%%%%%%%%%%%%%%%
\paragraph*{Logical Interpretation.}
%%%%%%%%%%%%%%%%%%%%%%%%%%%%%%

The third example is the degree of the plausibility of a given proposition
 conditioned by the prior information \cite{Jaynes03, Cox46, Cox61, Halpern99a, Halpern99b, vanHorn03}, 
 which is an extension of the truth value, and the central issue of this paper.
We hereafter show that the degree of the plausibility agrees with the conditions (\ref{Renyi}), indicating that 
the conditional probability is seen as a natural extension 
of the truth value of the proposition under uncertainty. 

In what follows, the Boolean operations play an important role,
since they characterize the relations among the (composite) propositions made 
through the logical operations. Given the truth values of the several propositions, 
we can find the truth values of the composite propositions associated with them by the Boolean operations. 
Since we will make the argument to see the conditional probability as an extension of the truth value, 
we shall provide a discussion on the Boolean algebra below.

First, let us show that the finitely additive class ${\cal F}=\{A, B, C\dots\}$ mentioned earlier is closed under the Boolean operations AND ($\land$), OR ($\lor$), and NOT ($\lnot$).
This implies that we may establish the logic on the finitely additive class, 
because the Boolean operation represents the relation among the logical propositions.

To this end, we introduce the Boolean operation AND, OR, NOT by
\be
A\land B:=A\cap B,
\quad
A\lor B:= A\cup B,
\quad
\lnot A:= A{}^c,
\ee
respectively.
We then clearly see the following properties, that is,
\begin{enumerate}
\item Idempotence
\be
&&A\land A=A\lor A=A\in{\cal F}.\label{idem}
\ee
\item Commutativity
\be
&&A\land B=B\land A\in{\cal F},\quad A\lor B=B\lor A\in{\cal F}.\label{comm}
\ee
\item Associativity
\begin{subequations}
\begin{align}
&A\land(B\land C)=(A\land B)\land C=A\land B\land C\in{\cal F},\label{assoc1}\\
&A\lor(B\lor C)=(A\lor B)\lor C=A\lor B\lor C\in{\cal F}.\label{assoc2}
\end{align}
\label{assoc}
\end{subequations}
\item Distributivity
\begin{subequations}
\begin{align}
&A\land(B\lor C)=(A\land B)\lor(A\land C)\in{\cal F},\label{dist1}\\
&A\lor(B\land C)=(A\lor B)\land(A\lor C)\in{\cal F}.\label{dist2}
\end{align}
\label{dist}
\end{subequations}
\item Duality
\begin{subequations}
\begin{align}
&C=A\land B\in{\cal F} \quad \Rightarrow \quad \lnot C=\lnot A\lor\lnot B\in{\cal F},\label{dual1}\\
&C=A\lor B\in{\cal F} \quad \Rightarrow \quad \lnot C=\lnot A\land\lnot B\in{\cal F}.\label{dual2}
\end{align}
\label{dual}
\end{subequations}
\end{enumerate}
Since these properties define the Boolean algebra, the finitely additive class is found to be closed under the Boolean operations.
On the basis of this observation, we may safely regard the element $\omega_i\in\Omega$ as an elementary proposition, which could be true or false.
The subsets $A, B, \dots\in{\cal F}$ are therefore the set of the composite propositions made by the Boolean operations, and the set $\Omega$ is seen as the set of all these.

Following the notation given in \cite{Cox46, Cox61, Kaynes21, Jaynes03, vanHorn03}, we then formally introduce the measure of conditional plausibility, denoted by $A|B\in\R$, whose value quantifies the degree of plausibility that
$A$ is true, given $B$ is true. 
We suppose that there exists two real numbers $\textit{\texttt{F}}$ and $\textit{\texttt{T}}$ such that $\textit{\texttt{F}} \le A|B\le \textit{\texttt{T}}$ for every $A$ and $B$. 
We hereafter show that $A|B$ satisfies the conditions (\ref{Renyi}) under reasonable assumptions given below.

Before we proceed, we point out that there are variants of the sets of the assumptions to prove Cox\rq s theorem \cite{Cox46, Cox61, Jaynes03, vanHorn03}. This is because Cox used implicit assumptions in his original proof \cite{Halpern99a, Halpern99b} and several researchers tried to construct more reasonable proof to fill the gap.
 Among these variants, we adopt the assumptions given in \cite{vanHorn03} for clarity and rigor of the presentation.

The first assumption is the following:
%%%%%%%%%%
\begin{assumption}[van Horn]
\label{dense}
There exists a non-empty set of real numbers $\texttt{P}_0$ with the following two properties:
% (i) $\texttt{P}_0$ is a dense subset of $(\texttt{F}, \texttt{T})$. (ii) For every $y_1, y_2, y_3\in\texttt{P}_0$, there exists propositions $A_1, A_2, A_3, B$ such that $y_1=A_1|B$, $y_2=A_2|A_1\land B$, and $y_3=A_3|A_2\land A_1\land B$.
\begin{enumerate}
  \item $\texttt{P}_0$ is a dense subset of the interval $(\texttt{F}, \texttt{T})$.
  \item For every $y_1, y_2, y_3\in\texttt{P}_0$, there exists propositions $A_1, A_2, A_3, B$ such that $y_1=A_1|B$, $y_2=A_2|A_1\land B$, and $y_3=A_3|A_2\land A_1\land B$.
\end{enumerate}
\end{assumption}
%%%%%%%%%%
Although the assumption \ref{dense} looks intricate, it is necessary to exclude the models which are not the standard probability theory, but satisfy all the  assumptions we will hereafter make:
Indeed, such a model is constructed in \cite{Halpern99a}. 
Moreover, it is known that the assumption \ref{dense} requires infinite numbers of the elementary propositions \cite{Halpern99b}.

%The first assumption is the following:
We make the second assumption:
%%%%%%%%%%
\begin{assumption}[van Horn]
\label{product}
Let $A, B, C$ be propositions. Then there exists a continuous function $F:[\texttt{F}, \texttt{T}]^2\rightarrow [\texttt{F}, \texttt{T}] $, which is strictly increasing in both the arguments, and satisfies
\be
A\land B|C=F(B|C,\,\, A|B\land C).\label{f}
\label{F}
\ee
\end{assumption}
%%%%%%%%f%%
The assumption \ref{product} implies that the plausibility of the composite proposition $A\land B$ given $C$ is true is related to two plausibilities $B|C$ and $A|B\land C$. In other words, we infer the plausibility of $A\land B|C$ from successive inferences with use of them. 

From the assumption \ref{dense} and \ref{product}, by using the properties of the Boolean algebra, it is found in \cite{vanHorn03} that the function $F(x,y)$ satisfies
\be
F(x, F(y,z))=F(F(x,y),z).
\label{ff}
\ee
Further, the following fact \cite{Aczel66} is useful to obtain the concrete expression of $F(x,y)$:
%%%%%%%%%%
\begin{lemma}[Acz\'el]
\label{associative_fnc}
Let $a$ and $b$, with $a<b$, be real numbers. Suppose that $f:[a,b]^2\rightarrow[a,b]$ is a continuous function, strictly increasing in both arguments, and satisfies the associativity equation
\be
f(x, f(y,z))=f(f(x,y),z)
\ee
for all $x, y, z \in(a,b]$. Then there exists some continuous, strictly increasing function $g(x)$ such that
\be
g(f(x,y))=g(x)+g(y).
\label{fg}
\ee
\end{lemma}
%%%%%%%%%%
Now we define $w(x):=e^{g(x)}$ and identify $f$ with $F$. By exponentiating both the sides of Eq.~(\ref{fg}) and substituting $w(x)$, we obtain
\be
F(x,y)=w^{-1}\[w(x)w(y)\],
%\quad
%0\le w(x)\le 1,
\label{w}
\ee
%where $w(x)$ is a continuous, strictly increasing, non-negative function to $[0,1]$. 
Note that substitution of Eq.~(\ref{w}) into Eq.~(\ref{F}) leads to
\be
w(A\land B|C)=w(B|C)w(A|B\land C),\label{w1}
\ee 
which is an expression of the inference rule (\ref{f}) in terms of $w(A|B)$. For the range of $w(x)$, we have $0\le w(x)\le 1$, whose proof is given in \cite{vanHorn03}.

Further, we make the following assumption on $w(x)$:
%%%%%%%%%%
\begin{assumption}[Cox]
\label{negation}
There exists a continuous, strictly decreasing, twice differentiable function $S(x)$ such that
\be
w(A|B)=S(w(\lnot A|B)).\label{s}
\ee 
\end{assumption}
%%%%%%%%%%
The assumption \ref{negation} means that the plausibility of a given proposition can be inferred from that of its negation. Cox found that Eq.~(\ref{s}) yields 
\be
S(x)=(1-x^m)^\frac{1}{m},
\ee
where $m$ is a positive finite constant.
Now we set $m=1$ to obtain
\be
w(A|B)+w(\lnot A|B)=1. \label{w2}
\ee

Eqs.~(\ref{w1}) and (\ref{w2}) suffice to show that the conditional plausibility
$w(A|B)$ satisfies Eqs.~(\ref{Renyi}). 
%Since Eq.~(\ref{null}) is shown to be valid by $0\le w(x)\le1$, what we have to prove are Eqs.~(\ref{omega}) and (\ref{ad}). 
To show these, we need
\be
w(A\lor B|C)=w(A|C)+w(B|C)-w(A\land B|C),%\nonumber\\
\label{sum}
\ee
whose proof is given in the Appendix.

We first prove Eq.~(\ref{abc}). Suppose $B\subset C$, which leads to $B\land C=B$. Then it follows from Eq.~(\ref{w1}) that
\be
w(A\land B|C)=w(A|B)w(B|C),
\label{w-red}
\ee
which reduces to Eq.~(\ref{abc}) when $w(B|C)\neq0$.

Furthermore, by setting $A=B=C$ in Eq.~(\ref{w-red}), we find $w(A|A)^2=w(A|A)$, whose solutions are either 1 or 0. Therefore we obtain Eq.~(\ref{one}) by taking the solution $w(A|A) =1$ and replacing $A$ with $B$.

Now we proceed to prove Eq.~(\ref{c-additivity}). Substituting $\Omega$ to $A$ in Eq.~(\ref{w1}), we find 
\be
w(\Omega\land B|C)=w(\Omega|C)w(B|\Omega\land C).\label{sub}
\ee
Since $\Omega\land B=B$ and $\Omega\land C=C$,
Eq.~(\ref{sub}) turns to
\be
w(\Omega|C)=1. \label{id}
\ee
%which proves Eq.~(\ref{omega}) for the fixed $C$.
Furthermore, putting $A=\Omega$ in Eq.~(\ref{w2}), we find
\be
w(\emptyset|C)=0.
\ee
Hence, from Eq.~(\ref{sum}), we obtain
\be
w(A\lor B|C) =w(A|C)+w(B|C)\quad \text{if}\quad A\land B=\emptyset.%\nonumber\\
\ee
By induction, we find
\be
w(\bigvee_{i=1}^n A_i|B) = \sum_{i=1}^n w(A_i|B),
\label{infsum}
\ee
for the set $\{A_i\}_{i=1}^n$, whose elements are mutually disjoint.
Note that the RHS of Eq.~(\ref{infsum}) converges, since the LHS is bounded.

%which shows Eq.~(\ref{c-additivity}). 
%By exchanging $C$ for $B$ and summing up all the results, we may set
The above argument guarantees that we may put
\be
\Pr(A|B)=w(A|B),
\ee
meaning that the degree of the plausibility can be seen as the conditional probability.

Note that we have not used the repeatability or frequency 
in the above argument. From the viewpoint of the logical interpretation, 
thus, we are allowed to apply the probability theory to not necessarily 
repeatable events.

%%%%%%%%%%%%%%%%%%%%%%%%%%%%%%
\section{Quantum Extension}
\label{extension}
%%%%%%%%%%%%%%%%%%%%%%%%%%%%%%
In the following, we extend the preceding argument
to quantum mechanics. Let ${\cal H}$ be a (countably) infinite dimensional Hilbert space, and
$P, Q, \cdots$ be the projectors acting on ${\cal H}$. 
The set of all the projectors is denoted by ${\cal L}({\cal H})$. The range of $P$ is denoted by $P({\cal H}):=\{ P\ket{\psi} \,|\, \ket{\psi}\in {\cal H}\}$. Note that $P(\cal H)$ is also the Hilbert space.
 
Since the eigenvalues of the projectors are either 1 or 0, the eigenvalues can be seen
as the truth values, and hence every projector can be thought of as a proposition.

Analogously to $A|B$ in the preceding section, let us now introduce $P|Q $ as the degree of the plausibility 
for the proposition $P$, given $Q$ is true. 
We suppose that $P|Q$ are bounded as $\textit{\texttt{F}}\le P|Q \le \textit{\texttt{T}}$ and well-defined for all the pairs of the projectors,
even for those which do not necessarily commute.

Given the propositions $P$ and $Q$, the composite projectors $P\land Q$ and $P\lor Q$ are defined as the projectors to the meet and join of  $P({\cal H})$ and $Q({\cal H})$, respectively. The negation $\lnot P$ is the projector on the orthogonal complement of $P({\cal H})$, which is hereafter denoted by $P({\cal H})^\bot $.

We shall show that the inference rule $P|Q$ obeys leads to the Born rule and the conditional probability satisfying the inference rule takes the form of the L\"uders rule.
For this pourpose, it is convenient to construct a Boolean subalgebra in a set of commuting projectors.
When $P$ and $Q$ commute with each other, we have the explicit expressions of the logical operations by associating them with corresponding projectors:
\be
P\land Q=PQ,
\quad
P\lor Q= P+Q-PQ,
\quad
\lnot P=\mathbbm{1}-P,%\nonumber\\
\ee
where $\mathbbm{1}$ is the identity operator. 
Note that $P\lor Q=P+Q$ if $P$ and $Q$ are mutually orthogonal.

Similarly to Eqs.~(\ref{defF}),
we can define a set ${\cal C}$ of commutative operators such that
\be
0&\in&{\cal C},\nonumber\\
P&\in&{\cal C} \quad \Rightarrow \quad \lnot P\in{\cal C},\nonumber\\
P, Q&\in&{\cal C} \quad \Rightarrow \quad P\land Q\in{\cal C}.\label{defC}
\ee
It is clear that the elements in ${\cal C}$ satisfies the properties 
(\ref{idem}), (\ref{comm}), (\ref{assoc}), (\ref{dist}), (\ref{dual}). Hence the set ${\cal C}$ is
found to be a Boolean subalgebra.

Three remarks are in order. First, $0$, $\mathbbm{1}$ in ${\cal C}$ correspond to $\emptyset$, $\Omega$ in ${\cal F}$, respectively.  Second, the set ${\cal C}$ is not unique: we can find infinitely many sets satisfying (\ref{defC}).
Indeed, given a projector $P$, we can construct ${\cal C}=\{0, P, \lnot P, \mathbbm{1}\}$, implying that any projector is an element of some Boolean subalgebra ${\cal C}$. Third, the join of all ${\cal C}$ gives ${\cal L}({\cal H})$.
 
We now extend the definitions of the probability and conditional probabiity given in Section~\ref{formulations} to quantum mechanics:
A non-negative function $\pro(P|Q)$ of the projectors $P\in{\cal L}({\cal H})$ and $Q\in{\cal K}$, where ${\cal K}$ is a non-empty subset of ${\cal L}({\cal H})$, is a conditional probability if it satisfies
 \begin{subequations}
%\begin{align}
\begin{align}
	\pro(Q|Q)&=1\quad\text{for}\quad \forall Q\in {\cal K}, 
	\label{qone}\\
	\pro(\bigvee_{i=1}^n P_i|Q)&= \sum_{i=1}^n\pro(P_i|Q) 	\label{qc-additivity}
\end{align}
for $P_i\in{\cal L}({\cal H})$ and $Q\in{\cal K}$, where $P_iP_j=P_i\delta_{ij}$ for any $i, j$, and
\begin{equation}
	\pro(P\land Q|Q)=\frac{\pro(P\land Q|R)}{\pro(Q|R)}
	\label{qabc}
\end{equation}
%\end{align}
for $P\in{\cal L}(\cal H)$ and $Q,R \in{\cal K} $ such that $Q< R$ and $\pro(Q|R)>0$. Here, $\delta_{ij}$ is the Kronecker delta, and $Q< R$ means that $Q({\cal H})\subset R({\cal H}) $, which leads to $Q=QR=RQ$ \cite{AG93}.
\label{qRenyi}
\end{subequations}
The probability is defined through the conditional probability as 
\be
\pro(P)=\pro(P|\mathbbm{1}),
\ee
which satisfies
\begin{subequations}
\begin{align}
\pro(\mathbbm{1})&=1, \\	
\pro(\bigvee_{i=1}^n P_i)&= \sum_{i=1}^n\pro(P_i), 
\end{align}	
for $P_i\in{\cal L}({\cal H})$ such that $P_iP_j=P_i\delta_{ij}$ for any $i, j$.
\end{subequations}

 The formal replacement of the sets $A,B,C$ to the projectors $P, Q, R$ in Eq.~(\ref{abc}) is insufficient to define the conditional probability in quantum mechanics, since such replacement overlooks the non-commutativity among the propositions. In other words, $\pro(P\land Q|Q)=\pro(P | Q)$ does not always hold in quantum mechanics, whereas its classical counterpart $\pro(A\land B|B)=\pro(A| B)$ always does by setting $B=C$ in Eq.~(\ref{abc}) \cite{Renyi56}. 
 
 The physical implication of $\pro(P\land Q|Q)\neq\pro(P | Q)$ is clear in the following example: let us take $P$ as the proposition that the momentum of a given particle has a definite value, say $p$, and take $Q$ as the other proposition that its position has a definite value $q$. On the one hand, then, $P\land Q$ is the proposition that the particle has the definite values of the position and momentum simultaneously, implying $\pro(P\land Q|Q)=0$ due to the uncertainty relation. On the other hand, $\pro(P|Q)$ may have the non-zero value, since the position eigenstate could yield the definite value $p$ in the momentum measurement.  
 
 Armed with this, we hereafter make the extension of Cox\rq s theorem.
 First, in parallel to the section \ref{interpretations}, we make three assumptions, which employ the analogues of $F$ and $S$ for commuting projectors. 
 %Although these may depend on the set ${\cal C}$, we do not suppose so. 
 Second, we derive the concrete expression of the conditional probability $\pro(P|Q)$ for the case $P\le Q$, from which the Born rule is derived.
 Here $P=Q$ stands for $P < Q$ and $Q < P$.
% Third, with the additional assumption, we show that the L\"uders rule satisfies Eqs.~(\ref{qRenyi}) for all the projectors.
 
The first assumption is as follows:
 %%%%%%%%%%
\begin{assumption}
\label{qdense}
There exists a non-empty set of real numbers $\texttt{P}_0$ with the following two properties:
% (i) $\texttt{P}_0$ is a dense subset of $(\texttt{F}, \texttt{T})$. (ii) For every $y_1, y_2, y_3\in\texttt{P}_0$, there exists propositions $A_1, A_2, A_3, B$ such that $y_1=A_1|B$, $y_2=A_2|A_1\land B$, and $y_3=A_3|A_2\land A_1\land B$.
\begin{enumerate}
  \item $\texttt{P}_0$ is a dense subset of $(\texttt{F}, \texttt{T})$.
  \item For every $y_1, y_2, y_3\in\texttt{P}_0$, there exist mutually commuting propositions $P_1, P_2, P_3, Q$ such that $y_1=P_1|Q$, $y_2=P_2|P_1\land Q$, and $y_3=P_3|P_2\land P_1\land Q$.
\end{enumerate}
\end{assumption}
%%%%%%%%%%
This assumption requires an infinite number of mutually commuting projectors in a Boolean subalgebra ${\cal C}$, excluding the finite dimensional Hilbert space.
%By making this assumption,  \textcolor{red}{\bf blah blah balh} in ${\cal C}$.
 
 To proceed, we make the following assumption:
%%%%%%%%%%
\begin{assumption}
\label{qproduct}
There exists a continuous function $G: [\texttt{F}, \texttt{T}]^2\rightarrow [\texttt{F}, \texttt{T}] $, which is strictly increasing in both the arguments, and satisfies
\be
P\land Q|R=G(Q|R,\,\, P|Q\land R) \label{g}
\ee
for any mutually commuting projectors $P,Q,R$.
\end{assumption}
%%%%%%%%%%
The function $G(x,y)$ ensures the validity of our inference on the composite proposition $P\land Q$ given $R$, if they commute. Conversely, if the propositions do not commute, then our inference does not necessarily hold, as shown in the section~\ref{violation}.

By using the assumptions \ref{qdense} and \ref{qproduct} and employing the argument given in the section~\ref{interpretations},
we can obtain
\be
G(G(x,y),z)=G(x, G(y,z)), \label{gg}
\ee
whose solution is given by
\be
G(x, y)=w^{-1}\[w(x)w(y)\],
\ee
where $w(x)$ is the continuous, strictly increasing none-negative function.
It follows that
\be
w(P\land Q|R)=w(Q|R)w(P|Q\land R)
\label{qw-red}
\ee
for any mutually commuting projectors $P,Q,R$.
Note that $0 \le w(P|Q)\le 1$ is also derived.

%Now we prove
%%%%%%%%%%
%\begin{lemma}
%\label{wpqr}
%There exists a continuous, strictly increasing, non-negative function $w(x)$ to $[0,1]$, which satisfies
%\be
%w(P\land Q|R)=w(Q|R)w(P|Q\land R)
%\label{qw-red}
%\ee
%for $P,Q,R\in{\cal C}$.
%\end{lemma}
%%%%%%%%%%
%\begin{proof}
%From Eq.~(\ref{g}), we may employ the discussion in the previous section in a given ${\cal C}$, and find a continuous, strictly increasing, non-negative function $w_{\cal C}(x)$ such that
%\be
%w _{\cal C}(P\land Q|R)=w _{\cal C}(Q|R)w _{\cal C}(P|Q\land R)
%\ee
%for $P,Q,R\in{\cal C}$. Then, given two Boolean subalgebra ${\cal C}$ and ${\cal C}^\prime$ having the non-vanishing intersection ${\cal C}\cap{\cal C}^\prime$, we find
%\be
%w_{\cal C}(P|Q)= w_{\cal C^\prime}(P|Q)
%\ee
%for $P,Q\in {\cal C}\cap{\cal C}^\prime $, suggesting that $w_{\cal C}(x) $ is independent of ${\cal C}$. Hence we may write $w_{\cal C}(x) $ as $w(x) $, which completes the proof.
%\end{proof}

% From the lemma~\ref{wpqr}, we immediately find
%%%%%%%%%%
%\begin{lemma}
%\label{proqq}
%The continuous, strictly increasing, non-negative function $w(x)$ to $[0,1]$ can be taken in a way that it satisfies Eq.~(\ref{qone}).
%\end{lemma}
%%%%%%%%%%
%\begin{proof}
%By setting $P=Q=R$ in Eq.~(\ref{qw-red}), we find $w(Q|Q)=0,1$.
%Thus Eq.~(\ref{qone}) is satisfied if we take the solution $w(Q|Q)=1$.
%\end{proof}

By finding the function $w(x)$, we safely put the next assumption:
%%%%%%%%%%
\begin{assumption}
\label{qnegation}
There exists a continuous, strictly decreasing, twice differentiable function $T(x)$ such that
\be
w(P|Q)=T(w(\lnot P|Q))\label{t}
\ee 
for any mutually commutative projectors $P,Q$.
\end{assumption}
%%%%%%%%%%
%The assumption \ref{qassum} means that our inference in the sense of Eq.~(\ref{f}) and (\ref{s}) is valid for any set ${\cal C}$.
This assumption leads to
\be
T(x)=1-x,
\label{T}
\ee
which reads
\be
w(P|Q)+w(\lnot P|Q)=1
\label{qw2}
\ee
for the projectors $P, Q$ such that $[P, Q]=0$.
%%%%%%%%%%
%\begin{lemma}
%The continuous, strictly decreasing function $T(x)$ can be taken as
%\be
%T(x)=1-x.
%\label{T}
%\ee
%\end{lemma}
%%%%%%%%%%
%\begin{proof}
% Similarly to the lemma~\ref{wpqr}, we utilize the argument in the previous section. Then, in a given ${\cal C}$, we obtain
% \be
% T(x)=(1-x^m)^\frac{1}{m},
% \ee
% where $m$, being a positive finite constant, may be dependent of ${\cal C}$.\\
% We now suppose that we are given two Boolean subalgebra ${\cal C}$ and ${\cal C}^\prime$ having the non-vanishing intersection ${\cal C}\cap{\cal C}^\prime$. On this intersection, we find
%\be
%[1-w(\lnot P|Q)^m]^\frac{1}{m}= [1-w(\lnot P|Q)^{m^\prime}]^\frac{1}{m^\prime},
%\ee
%for $P,Q\in {\cal C}\cap{\cal C}^\prime $, suggesting that $m=m^\prime$. By setting $m=1$, we arrive at Eq.~(\ref{T}).
%\end{proof}
Hence it follows from Eqs.~(\ref{qw-red}) and (\ref{qw2}) that
\be
w(P\lor Q|R)=w(P|R)+w(Q|R),%\nonumber\\
\label{gl}
\ee
for any mutually commuting projectors $P,Q,R$. 
%This shows that $w(P|Q)$ is a probability measure for every ${\cal C}$, if $R$ is fixed.
%Moreover, by fixing $Q$ and taking into account all the sets ${\cal C}$ having $Q$ as its element, we easily find that Eq.~(\ref{qc-additivity}) holds for all $P_i$ such that $P_iP_j=\delta_{ij}P_i$ for the fixed $Q$. Note that $[P_i, P_j]=0$.

We now proceed to prove a technical lemma, which plays an essential role to determine the explicit form of $w(x)$ and show the uniqueness of the (conditional) probability measure associated with $w(x)$ in quantum mechanics.
%%%%%%%%%%
\begin{lemma}
\label{uniqness}
Suppose that given a projector $Q$ and two density operators $\rho_1$ and $\rho_2$ such that
\be
\tr(\rho_1Q)= \tr(\rho_2Q)=1,
\label{r1r2Q}
\ee
and
\be
\tr(\rho_1P)= \tr(\rho_2P),
\label{r1r2P}
\ee
for all $P$ such that $P\le Q$.
Then, 
\be
\rho_1=\rho_2.
\ee
\end{lemma}
%%%%%%%%%%
\begin{proof}
From Eq.~(\ref{r1r2Q}), we obtain
\be
\tr(\rho_1\lnot Q)= \tr(\rho_2\lnot Q)=0.
\label{r1r2not}
\ee
Let us perform the spectral decomposition to write $\lnot Q$ as
\be
\lnot Q = \ket{\varphi}\bra{\varphi}+\dots.
\label{qspectrum}
\ee
By substituting Eq.~(\ref{qspectrum}) to Eq.~(\ref{r1r2not}), we find
\be
\tr(\rho_i\lnot Q)=\tr(\rho_i \ket{\varphi}\bra{\varphi})+\dots=0,
\ee
for $i=1,2$.
Since density operators and projectors are non-negative, we find $\tr(\rho_i \ket{\varphi}\bra{\varphi}) =\bra{\varphi}\rho_i\ket{\varphi}=0$. This implies that
\be
\rho_i\ket{\varphi}=0,
\label{kernel}
\ee 
for any vector $\ket{\varphi}\in Q({\cal H})^\bot$.

Let us take a normalized vector $\ket{\psi}\in{\cal H}$ and decompose it as
\be
\ket{\psi} = \ket{\psi_1}+\ket{\psi_2},
\ee
where
\be
\ket{\psi_1}=Q\ket{\psi}\in Q({\cal H}),
\quad
\ket{\psi_2}=(\mathbbm{1}-Q)\ket{\psi}\in Q({\cal H})^\bot.%\nonumber\\
\ee
Since Eq.~(\ref{kernel}) holds for $\ket{\psi_2}\in Q({\cal H})^\bot $,  for a projector $R= \ket{\psi}\bra{\psi} $, we observe
\be
\tr(\rho_i R)
&=&\bra{\psi}\rho_i\ket{\psi}\nonumber\\
&=& \bra{\psi_1}\rho_i\ket{\psi_1}+ \bra{\psi_2}\rho_i\ket{\psi_2}+ 2{\rm Re}\bra{\psi_1}\rho_i\ket{\psi_2}\nonumber\\
&=& \bra{\psi_1}\rho_i\ket{\psi_1}\nonumber\\
&=& \|\psi_1\|^2\tr(\rho_i O)
\ee
for $i=1,2$. Here $\|\psi_1\|=\sqrt{\braket{\psi_1 |\psi_1}}$ is the norm of the vector $\ket{\psi_1}$ and $O =\ket{\psi_1}\bra{\psi_1}/\|\psi_1\|^2$ is the projector onto the one-dimensional subspace generated by $\ket{\psi_1}$.
%, that is, $O=\ket{\psi_1}\bra{\psi_1}/\|\psi_1\|^2$.

With the use of $\ket{\psi_1}\in Q({\cal H})$, we find $O\le Q$.
Then it follows from Eq.~(\ref{r1r2Q}) that
\be
\tr(\rho_1 O)=\tr(\rho_2 O),
\ee
which leads to
$
\tr(\rho_1 R)=\tr(\rho_2 R),
$
or equivalently,
\be
\tr(DR)=0,
\label{dr}
\ee
where
\be
D=\rho_1-\rho_2.
\ee
Since the operator $D$ is normal ($DD^\dagger=D^\dagger D$, where $\dagger$ stands for the adjoint operation) and Eq.~(\ref{dr}) holds for any one-dimensional projector $R$, which is clearly non-negative, we can employ the spectral decomposition and obtain $\rho_1=\rho_2$, which completes the proof.
\end{proof}

The lemma \ref{uniqness} is a special case of the lemma~\ref{BC_lemma}, which is first shown in \cite{BC81, Malley04}.

%We hereafter generalize this observation:
We prove the following lemma.
%%%%%%%%%%
\begin{lemma}
\label{expression}
If $\dim Q({\cal H})\ge3$, there exists a density operator $\rho$,
with which $w(P|Q)$ is written as
\be
w(P|Q)=\frac{\tr(\rho P)}{\tr(\rho Q)}
\label{pre_lueders}
\ee
for all $P$ such that $P\le Q$.
\end{lemma}
%%%%%%%%%%
\begin{proof}
By using induction to Eq.~(\ref{gl}), we obtain
\be
w(\bigvee_{i=1}^n P_i|Q) = \sum_{i=1}^n w(P_i|Q),
\label{qinfsum}
\ee
%It is straightforward to find Eq.~(\ref{qc-additivity}) by induction,
for the projectors $P_1,\dots, P_n, Q$ such that $P_iP_j=\delta_{ij}P_i$ and $[P_i, Q]=0$. 
Since $[P_i,Q]=0$ is the necessary condition of $P_i\le Q$,
Eq.~(\ref{qinfsum}) also holds for the set of mutually orthogonal projectors $\{P_i\}_{i=1}^n$ such that $P_iP_j=\delta_{ij}P_i $ and $P_i\le Q$ for the fixed $Q$. Furthermore, by noticing that $Q$ behaves like the identity operator on $Q(\cal H)$ due to the idempotence $Q^2=Q$, we observe that Eq.~(\ref{qinfsum}) can be seen as the sufficient condition of the Gleason theorem \cite{Gleason57}, as far as we consider the set of the projectors $\{P\,|\, P\le Q\}$.

On the basis of the above argument, we may employ the Gleason theorem,
and hence obtain
\be
w(P|Q)=\tr(\rho_QP).\label{pq}
\label{wpq}
\ee
for $P\le Q$, if $\dim Q({\cal H})\ge3$. Here $\rho_Q$ is a density operator which is dependent on $Q$. 
Since $Q\le Q$, we can set $P=Q$ to observe that
\be
w(Q|Q)=\tr(\rho_QQ)=1
\label{wqq}
\ee
from Eq.~(\ref{qone}). 
The density operator $\rho_Q$ satisfying Eq.~(\ref{wpq}) and Eq.~(\ref{wqq}) for $\{P\,|\, P\le Q\}$ is uniquely determined, due to the lemma \ref{uniqness}.

Now it is clear from Eq.~(\ref{wqq}) that the density operator $\rho_Q$ has the non-trivial support only on the eigenspace of $Q$. Hence, with an appropriate density operator $\rho$, we can write
\be
\rho_Q=\frac{Q\rho Q}{\tr(\rho Q)}.\label{rhoq}
\ee
Plugging Eq.~(\ref{rhoq}) into Eq.~(\ref{pq}) and using the cyclic property of trace and $PQ=QP=P$ coming from $P\le Q$, we find Eq.~(\ref{pre_lueders}), which completes the proof.
\end{proof}

%%%%%%%%%%
\begin{lemma}
\label{pre_luders_cp}
$w(P|Q)$ is the conditional probability for $P, Q$ such that $P\le Q$.
\end{lemma}
%%%%%%%%%%
\begin{proof}
To prove this, we show Eqs.~(\ref{qRenyi}) for Eq.~(\ref{pre_lueders}).
Eq.~(\ref{qone}) and (\ref{qc-additivity}) are obvious.
For Eq.~(\ref{qabc}), we find
\be
\frac{w(P\land Q|R)}{w(Q|R)}=\frac{\tr[\rho(P\land Q)]}{\tr(\rho Q)}=w(P\land Q |Q),
\label{pre_trpqr}
\ee
which completes the proof.
\end{proof}

It is now straightforward to derive the Born rule:
%%%%%%%%%%
\begin{theorem}
The probability to find the proposition $P$ being true is given 
by the Born rule:
\be
\Pr(P)=\tr(\rho P).
\label{Born}
\ee
\end{theorem}
%%%%%%%%%%
\begin{proof}
We set $Q=\mathbbm{1}$ in Eq.~(\ref{pre_lueders}). Then we can make use of the lemma \ref{expression} and obtain
\be
w(P|\mathbbm{1})=\tr(\rho P).
\ee
%It is clear that $w(\mathbbm{1} |\mathbbm{1})=1$ and $w(\bigvee_{i=1}^\infty P_i |\mathbbm{1})= \sum_{i=1}^\infty w(P_i|\mathbbm{1})$ for $P_i$ such that $P_iP_j=\delta_{ij}P_i$, $w(\cdot|\mathbbm{1})$ is seen as the probability. Thus from Eqs.~() 
Since $w(P|Q)$ is the conditional probability for $P\le Q$ and $P\le\mathbbm{1}$ holds for any projector $P$, we may set Eq.~(\ref{Born}), which proves the theorem.
\end{proof}

Since the identity operator $\mathbbm{1}$ corresponds to $\Omega$, this corollary shows that we can regard the Born rule as the degree of the plausibility of the proposition $P$ under no information.

%%%%%%%%%%%%%%%%%%%%%%%%%%%%%%
\section{Inference Rule and L\"uders Rule}
\label{Lueders}
%%%%%%%%%%%%%%%%%%%%%%%%%%%%%%

Equation (\ref{pre_lueders}) is seen as the special case of the L\"uders rule of the conditional probability \cite{Busch09}, which takes the form of
\be
\pro(P|Q)=\frac{\tr(Q\rho QP)}{\tr(\rho Q)}.
\label{lueders}
\ee
Indeed, when $P\le Q$, the numerator of the RHS is written as $\tr(Q\rho QP)= \tr(Q\rho P)= \tr(\rho PQ)= \tr(\rho P)$, which results in Eq.~(\ref{pre_lueders}).
%even though its domain is restricted by the projector ordering $P\le Q$. Moreover, Eqs.~(\ref{qc-additivity}) and (\ref{qabc}) has not yet been proven.
%We hereafter relax the restriction on the projectors and show that the conditional probability takes the form of Eq.~(\ref{lueders}) even for the arbitrary pairs of the projectors $P$ and $Q$.
%fulfills Eqs.~(\ref{qc-additivity}) and (\ref{qabc}) as follows:
We now turn to show that the L\"uders rule is the concrete expression of the conditional probability which is defined for arbitrary pairs of the projectors and in accordance with logical interpretation.
To this end, we recall
%%%%%%%%%%
\begin{lemma}[Beltrametti and Cassinelli \cite{BC81, Malley04}]
\label{BC_lemma}
For Hilbert space ${\cal H}$, let $\pro(\cdot)$ be a probability measure on ${\cal L}({\cal H})$, and let $Q$ be any projector such that $\pro(Q	)\neq0$. Then there exists a unique probability measure on ${\cal L}({\cal H})$, which we denote by
$\pro(\cdot| Q)$, such that for all projectors $P\le Q$ it is the case that $\pro(P|Q) =\pro(P)/\pro(Q)$.
\end{lemma}
%%%%%%%%%%
\begin{proof}
From the Gleason theorem, we find that, for any projector $Q$, 
the probability measure $\pro(\cdot| Q) $ must take the form $\pro(P|Q)=\tr(\rho_QP)$ for some density operator $\rho_Q$.
If we can show the uniqueness of the density operator $\rho_Q$, then we complete the proof. 

Now suppose that 
$\rho_1$ and $\rho_2$ are two density operators such that
$\tr(\rho_1P) = \tr (\rho_2P) =\pro(P)/\pro(Q)$
for all projectors $P$, which satisfies $P\le Q$.
We then find Eq.~(\ref{r1r2Q}) by setting $P=Q$.
Therefore, we can employ the lemma \ref{uniqness} and obtain $\rho_1=\rho_2$, which completes the proof.
\end{proof}

%Motivated by the lemma \ref{BC_lemma}, we make the following assumption.
%%%%%%%%%%
%\begin{assumption}
%\label{apriori}
%The degree of plausibility $w(P|Q)$ is a conditional probability for any pairs of non-commuting projectors $P, Q \in {\cal L}({\cal H})$, that is, $\pro(\cdot | Q)=w(\cdot | Q)$ for any projector $P\in{\cal L}({\cal H})$.
%\end{assumption}
%%%%%%%%%%

%Summing up all the lemmata, we arrive at
We arrive at
%%%%%%%%%%
\begin{theorem}
If there exists a conditional probability satisfying
\be
\Pr(P\land Q|R)=\Pr(P|R)\Pr(Q|P\land R)
\label{inference}
\ee
for any commuting projectors $P, Q, R$, then it is uniquely determined and takes the form of Eq.~(\ref{lueders}).
%If the assumption~\ref{qdense}, \ref{qproduct}, \ref{qnegation} and \ref{apriori} hold, then there exists the unique conditional probability, which takes the form of Eq.~(\ref{lueders}).
\end{theorem}
%%%%%%%%%%
\begin{proof}
We first show that Eq.~(\ref{inference}) leads to $\pro(P|Q) =\pro(P)/\pro(Q)$ for $P\le Q$.
Since $P\le\mathbbm{1}$ and $\pro(P)=\pro(P|\mathbbm{1})$, we make use of Eq.~(\ref{inference}) to obtain   
\be
\pro(Q)\pro(P|Q)= \pro(Q|\mathbbm{1})\pro(P|Q\land\mathbbm{1})= \pro(P\land Q|\mathbbm{1})=\pro(P\land Q)=\pro(P).
\ee
Here we have used $P\le Q$ to the last equation.

%From the lemma~\ref{expression} and assumption~\ref{apriori}, 
Next, let us use the lemma~\ref{BC_lemma}. Then we find that there exist a unique density operator $\rho_Q$, by which the conditional probability takes the form $\pro(P|Q)=\tr(\rho_Q P)$. By setting $P=Q$, then we obtain $\pro(Q|Q)=1$.
Thus, as with the proof of the lemma~\ref{expression}, we find that there exists a density operator $\rho$ such that $\rho_Q=Q\rho Q/\tr(\rho Q)$.
%Furthermore, when $P\le Q$, we have
%$
%\pro(P|Q)=\tr(\rho P)/\tr(\rho Q)
%$
%from Eq.~(\ref{pre_lueders}).
%By the uniqueness of the density operator, we can take $\sigma=\rho$, 
which completes the proof.
\end{proof}

Thus we have arrived at the L\"uders rule as the conditional probability for all the projectors, by starting from the inference rules valid at least in mutually commuting projection operators. 
%Since the inference rule is at the heart of the logical interpretation,  once we admit the apriori hypotheses that the degree of plausibility is a conditional probability for the non-commuting projectors.

In the rest of this section, we show that the L\"uders rule satisfies Eqs.~(\ref{qRenyi}).
%%%%%%%%%%
\begin{corollary}
\label{additivity-lemma}
The equation (\ref{lueders}) satisfies Eq.~(\ref{qc-additivity}) for $P_i\in{\cal L}({\cal H})$ and $Q\in{\cal K}$ such that $P_iP_j=\delta_{ij}P_i$.
\end{corollary}
%%%%%%%%%%
\begin{proof}
This is shown by the direct calculation.
\end{proof}

%%%%%%%%%%
\begin{corollary}
\label{consistency}
The equation (\ref{lueders}) satisfies Eq.~(\ref{qabc}).
\end{corollary}
%%%%%%%%%%
\begin{proof}
From Eq.~(\ref{lueders}), we observe
\be
\frac{\pro(P\land Q|R)}{\pro(Q|R)}=\frac{\tr[(P\land Q)R\rho R]}{\tr(QR\rho R)}.
\label{trpqr}
\ee
Since $Q<R$, we have $Q=QR=RQ$, which reduces the denominator of the RHS of Eq.~(\ref{trpqr}) to
\be
\tr(QR\rho R)= \tr(Q\rho R) = \tr(\rho RQ) = \tr(\rho Q). 
\label{deno}
\ee
Similarly, we have $P\land Q<Q$, since $(P\land Q)({\cal H})\subset Q({\cal H})$. Thus, by the same fashion as Eq.~(\ref{deno}), we obtain
\be
\tr[(P\land Q)R\rho R]&=& \tr[(P\land Q) QR\rho R]\nonumber\\
& = &\tr[(P\land Q)Q\rho R]\nonumber\\
& = &\tr[Q\rho R (P\land Q)]\nonumber\\
& = &\tr[Q\rho RQ (P\land Q)]\nonumber\\
& = &\tr[Q\rho Q (P\land Q)].
\label{nume}
\ee
By plugging Eq.~(\ref{deno}) and (\ref{nume}) into Eq.~(\ref{trpqr}),
we obtain Eq.~(\ref{qabc}), which completes the proof.
\end{proof}

%%%%%%%%%%%%%%%%%%%%%%%%%%%%%%
\section{Violation of the Inference Rule}
\label{violation}
%%%%%%%%%%%%%%%%%%%%%%%%%%%%%%
In the previous section, we have derived the L\"uders rule 
by assuming the validity of the inference rules (\ref{trpqr}) for the projectors $P, Q$ with the ordering relation $P\le Q$. This suggests that the inference rules
do not necessarily hold for non-commuting projectors.
 
However, the relation between the validity of the inference rule and the commutativity is not simple. Indeed, as suggested in the following proposition, we can show that the inference rule holds even when there exists {\it a} pair of non-commuting projectors. 
To argue this, we introduce 
\be
\Delta=\Pr(P\land Q|R)-\Pr(P|R)\Pr(Q|P\land R).
\ee
and check its behavior.
If $\Delta=0$, then the inference rule (\ref{w1}) holds.

We first show
\begin{proposition}
Suppose $[P,Q]=[P,R]=0$	. Then $\Delta=0$.
\end{proposition}
\begin{proof}
We show $\Pr(P\land Q|R)=\Pr(P|R)\Pr(Q|P\land R)$.
Since $[P,Q]=[P,R]=0$, we have
\be
\Pr(P\land Q|R)&=&\frac{\tr[R\rho R(P\land Q)]}{\tr(\rho R)}
= \frac{\tr(R\rho RPQ)}{\tr(\rho R)},\nonumber\\
\Pr(P|R)&=&\frac{\tr(\rho RP)}{\tr(\rho R)},\nonumber\\
\Pr(Q|P\land R)&=&\frac{\tr[(P\land R)\rho (P\land R)Q]}{\tr(\rho P\land R)}
= \frac{\tr(PR\rho PRQ)}{\tr(\rho PR)}
= \frac{\tr(R\rho RPQ)}{\tr(\rho RP)},
\ee
by using the relations $P^2=P$, $Q^2=Q$, $R^2=R$ and the ciclic property of the trace $\tr{(PQ)}=\tr{(QP)}$.
It thus follows that $\Pr(P\land Q|R)=\Pr(P|R)\Pr(Q|P\land R)$, which completes the proof.
\end{proof}

This proposition suggests that $[Q,R]\neq0$ is insufficient for $\Delta\neq0$: The violation of the inference rule requires either $[P,Q]\neq0$ or $[P,R]\neq0$.
 
Second, we turn to construct an example of the violation of the inference rule (\ref{w1}).
%\begin{example}
Now we set ${\cal H}=\C^3 $ and introduce the normalized state vectors
\be
\vpsi &=&(\psi_1, \psi_2, \psi_3)^t, \quad (\psi_i\in\R) \nonumber\\ 
\vphi &=&(\varphi_1, \varphi_2, \varphi_3)^t, \quad (\varphi_i\in\R) \nonumber \\
\vz &=&(0, 0, 1)^t,
\ee
where $t$ is the transposition operation.
We further define the projector onto the two-dimensional planes as
\be
P=\mathbbm{1}_3-\vpsi\vpsi^t, \quad
Q=\mathbbm{1}_3-\vphi\vphi^t, \quad
R=\mathbbm{1}_3-\vz\vz^t,%\nonumber \\
\ee
where $\mathbbm{1}_3$ stands for the identity operator on $\C^3$.

The meet $P\land Q$ is the projector onto the intersection of the two 2-dimensional subspaces $P({\cal H})$ and $Q({\cal H})$, which are seen as planes. Thus, if $P\neq Q$, $P\land Q$ is the projector
onto the vector orthogonal to the normal vectors $\vpsi$ and $\vphi$:
\be
P\land Q=\vxi(\vpsi,\vphi) \vxi(\vpsi,\vphi)^t,
\ee
where
\be
\vxi(\vpsi,\vphi) =\frac{\vpsi\times\vphi}{|\vpsi\times\vphi|},
\ee
and $\times$ is the exterior product.
Note that
\be
P \vxi(\vpsi,\vphi)=Q\vxi(\vpsi,\vphi)= \vxi(\vpsi,\vphi),
\ee
which ensures $\vxi(\vpsi,\vphi) \in P\land Q({\cal H})$.
By the same argument, we observe
\be
P\land R=\vxi(\vpsi,\vz) \vxi(\vpsi,\vz)^t,
\ee
when $P\neq R$.

Furthermore, let us set $\rho=\mathbbm{1}_3/3$. By the direct calculation, we then obtain
\be
\Pr(P \land Q|R)&=&\frac{1}{2}\(1-\frac{|(\vpsi\times\vphi)\cdot\vz|^2}{|\vpsi\times\vphi|^2}\), \nonumber\\
\Pr(P|R)&=&\frac{1}{2}\(1+|\vpsi\cdot\vz|^2\), \nonumber\\
\Pr(Q|P \land R)&=&1-\frac{|(\vpsi\times\vz)\cdot\vphi|^2}{|\vpsi\times\vz |^2},
\ee
which lead to
\be
\Delta &=&\frac{1}{2}\biggl[1-\frac{|(\vpsi\times\vphi)\cdot\vz|^2}{|\vpsi\times\vphi|^2}
-\(1+|\vpsi\cdot\vz|^2\)\(1-\frac{|(\vpsi\times\vz)\cdot\vphi|^2}{|\vpsi\times\vz |^2}\)\biggr],
\ee
for $P\neq Q\neq R$.

%Now we set ${\cal H}=\C^{2\otimes2}$ and
%\be
%&&P=\ket{\uparrow}\bra{\uparrow}\otimes\mathbb{1}_2,
%\quad
%Q=\mathbb{1}_2\otimes\ket{\uparrow}\bra{\uparrow},\nonumber\\
%&&R=\mathbb{1}_2\otimes\ket{\rightarrow}\bra{\rightarrow},
%\ee
%where $\mathbb{1}_2$ stands for the identity operator on $\C^2$.
%We have defined 
%$\ket{\uparrow}=\(\begin{smallmatrix}
%      1 \\
%      0 
%    \end{smallmatrix}\)$,
%$\ket{\downarrow}=\(\begin{smallmatrix}
%      0 \\
%      1 
%    \end{smallmatrix}\)$,
%and
%$\ket{\rightarrow}=\frac{1}{\sqrt{2}}\(\begin{smallmatrix}
%      1 \\
%      1 
%    \end{smallmatrix}\)$.
% Note that $[P,Q]=[P,R]=0$, but $[Q,R]\neq0$.
 
%Furthermore, let us set
%\be
%\rho=\frac{r}{4}\mathbb{1}_2^{\otimes2}+(1-r)\ket{\uparrow}\bra{\uparrow}^{\otimes2},
%\quad
%0\le r\le1,
%\ee
%which is a mixture of the completely mixed state and a pure separable state.
%We then find
%\be
%\Delta=\frac{\sqrt{2}-1}{2}(r-1).
%\ee

We now find the violation of the inference rule (\ref{w1}):
when $\vpsi=(1/\sqrt{2})(1,0,1)^t$ and $\vphi= (1/\sqrt{2})(0,1,1)^t$, we have $\Pr(P\land Q|R)=1/3$, $\Pr(P|R)=3/4$, and $\Pr(Q|P \land R)=1/2$, resulting in $\Delta=-1/24$. Note that $P,Q,R$ are mutually non-commuting in this case.
%\end{example}

%%%%%%%%%%%%%%%%%%%%%%%%%%%%%%
\section{Conclusion and Discussions}
\label{conclusion}
%%%%%%%%%%%%%%%%%%%%%%%%%%%%%%
We obtained the Born rule and the L\"uders rule through the simple and natural
generalization of the logical interpretation of the probability theory.
 We need not think of both the rules as postulates, but as theorems deduced from the reasonable 
assumption that our inference makes sense for any set of the commuting 
projectors.

Whereas the inference rules hold for any $\cal{C}$ by the assumptions,
it does not hold for non-commuting projectors. This property of the L\"uders
rule is analogous to the algebraic property of the observables appearing
in Mermin\rq s magic square \cite{Mermin93}: 
Given a set of non-commuting observables,
subsets of the commuting observables follow the algebraic relations of their
eigenvalues, but it is not the case for the total set of the non-commuting observables. We could say that the validity of our inference rules depends
on the context. This property has been overlooked in \cite{Holik14}, where the Born rule has been algebraically derived by using the associativity relation of the composite proposition $(P\lor Q)\lor R = P\lor (Q\lor R) =P\lor Q\lor R$ for mutually orthogonal projectors $P,Q,R$.

The results we have obtained could give new insights to quantum logic \cite{Pitowsky89}.
In the quantum logic, we can show that (i) the projectors of the Hilbert space are seen as representations
of elements of the orthomodular lattice, which governs the algebraic relations among
quantum propositions \cite{Birkhoff36}, and (ii) the probability measure on the projectors
is given by the Born rule (the Gleason theorem). Note that these outcomes
are formal, and suggest nothing about the interpretations of the quantum probability.

In contrast, we have derived the Born rule on the assumptions which are easy to understand 
as the  inference rules in the Boolean subalgebra.
Thus, we may say that, on the basis of the quantum logic, we have arrived at an interesting and new viewpoint: the Born rule is the natural 
extension of the standard inference rules to the orthomodular lattice structure.

We mention that the inference approach is one of the conceptual 
foundations in machine learning \cite{Bishop06}. In particular, various non-informative
prior distributions such as Jeffery's prior are proposed on the line of thought of the logical interpretation. Analogously, our results obtained here may provide a conceptual basis of quantum machine learning \cite{Biamonte17}, which is a recent central issue in quantum information theory \cite{Preskill18}.

%\acknowledgments
\section*{Acknowledgment}
We thank I. Tsutsui and S. Hanashiro for fruitful discussions.

\appendix
\section{Proof of Eq. (\ref{sum})}
To prove Eq.~(\ref{sum}), we follow the argument given in \cite{Jaynes03}.
By using Eqs. (\ref{dual2}), (\ref{w1}) and (\ref{s}), we obtain
\be
w(A\lor B|C)&=&1-w(\lnot (A\lor B)|C) \nonumber\\
&=&1-w(\lnot A\land\lnot B|C) \nonumber\\
&=&1-w(\lnot A|C)w(\lnot B|\lnot A\land C) %\quad \text{from Eq.~(\ref{w1})} 
\nonumber\\
&=&1-w(\lnot A|C)\[1-w(B|\lnot A\land C)\] \nonumber\\
&=&w(A|C)+w(\lnot A|C)w(B|\lnot A\land C) %\quad %\text{from Eq.~(\ref{s})}
\nonumber\\
&=&w(A|C)+w(\lnot A\land B|C) %\quad \text{from Eq.~(\ref{w1})}
\nonumber\\
&=&w(A|C)+w(B|C)w(\lnot A|B\land C) %\quad \text{from Eq.~(\ref{w1})}
\nonumber\\
&=&w(A|C)+w(B|C)\[1-w(A|B\land C)\] \nonumber\\
&=&w(A|C)+w(B|C)-w(A\land B|C),
\ee
which completes the proof.

%--------------------------%

\end{document}